\documentclass[proof]{WileyASNA-v1CD}

\usepackage{graphicx}

\usepackage{amsmath}
\usepackage{hyperref}
\usepackage{natbib}
\usepackage[rightcaption]{sidecap}
 \usepackage[english]{babel}

\hypersetup{
    final=true,
    pageanchor=true,
    colorlinks=true,
    breaklinks=true,
    linkcolor=blue,
    citecolor=blue,
    urlcolor=blue,
    pdfpagemode=UseNone,
    pdftitle={},
    pdfauthor={Shen},
    pdfsubject={Solar Physics},
    pdfkeywords={}}

\graphicspath{{figures/}}

\newcommand\arcsec{\mbox{$^{\prime\prime}$}}

\newcommand{\monthword}[1]{\ifcase#1\or January\or February\or March\or April\or
                                        May\or June\or July\or August\or
                                        September\or October\or November\or 
                                        December\fi}
\newcommand{\todayI}{\the\day~\monthword{\the\month}~\the\year}      

\newcommand{\ns}{\hspace*{-5pt}}

\articletype{Original Article}%

\received{\todayI}
\revised{DD Month YYYY}
\accepted{19 November 2018}

\begin{document}

\title{Calibration of full-disk \mbox{He\,\textsc{i}} 10\,830~\AA\ 
filtergrams of the Chromospheric Telescope}

\author[1,2]{Zili Shen}

\author[1,3]{Andrea Diercke*}

\author[1]{Carsten Denker}

\authormark{SHEN \textsc{et al}}

\address[1]{\orgname{Leibniz-Institut f\"ur Astrophysik Potsdam (AIP)}, 
\orgaddress{\state{Potsdam}, \country{Germany}}}

\address[2]{\orgdiv{Department of Astronomy}, \orgname{University of Texas at 
Austin}, \orgaddress{\state{Texas}, \country{USA}}}

\address[3]{\orgdiv{Institut f\"ur Physik und Astronomie}, 
\orgname{Universit\"at Potsdam}, \orgaddress{\state{Potsdam}, 
\country{Germany}}}

\corres{*Leibniz-Institut f\"ur Astrophysik Potsdam (AIP), An der Sternwarte 
16, 14482 Potsdam, Germany, \email{adiercke@aip.de}}


\abstract{The Chromospheric Telescope (ChroTel) is a small 10-cm robotic 
telescope at Observatorio del Teide on Tenerife (Spain), which observes the 
entire Sun in H$\alpha$, \mbox{Ca\,\textsc{ii}\,K}, and \mbox{He\,\textsc{i}} 
10\,830~\AA. We present a new calibration method that includes limb-darkening 
correction, removal of non-uniform filter transmission, and determination of 
\mbox{He\,\textsc{i}} Doppler velocities.

Chromospheric full-disk filtergrams are often obtained with Lyot filters, which 
may display non-uniform transmission causing large-scale intensity variations 
across the solar disk. Removal of a 2D symmetric limb-darkening function from 
full-disk images results in a flat background. However, transmission artifacts 
remain and are even more distinct in these contrast-enhanced images. Zernike 
polynomials are uniquely appropriate to fit these large-scale intensity 
variations of the background. The Zernike coefficients show a distinct temporal 
evolution for ChroTel data, which is likely related to the telescope's 
alt-azimuth mount that introduces image rotation. In addition, applying this 
calibration to sets of seven filtergrams that cover the \mbox{He\,\textsc{i}} 
triplet facilitates determining chromospheric Doppler velocities. To validate 
the method, we use three data sets with varying levels of solar activity. The 
Doppler velocities are benchmarked with respect to co-temporal high-resolution 
spectroscopic data of the GREGOR Infrared Spectrograph (GRIS). Furthermore, 
this technique can be applied to ChroTel H$\alpha$ and \mbox{Ca\,\textsc{ii}}~K 
data. The calibration method for ChroTel filtergrams can be easily adapted to 
other full-disk data exhibiting unwanted large-scale variations. The spectral 
region of the \mbox{He\,\textsc{i}} triplet is a primary choice for 
high-resolution near-infrared spectropolarimetry. Here, the improved calibration 
of ChroTel data will provide valuable context data.}


\jnlcitation{\cname{%
\author{Z. Shen}, 
\author{A. Diercke}, and 
\author{C. Denker}} (\cyear{2018}), 
\ctitle{Calibration of full-disk \mbox{He\,\textsc{i}} 10\,830~\AA\ 
filtergrams of the Chromospheric Telescope}, \cjournal{Astron. Nachr./AN.}, 
\cvol{????}.}


\maketitle



\section{Introduction}

The Sun has been observed with telescopes for more than 400~years on a regular 
basis. Long-term observations of sunspots revealed a cyclic behavior of their 
appearance related to the solar dynamo \citep{Cliver2014}. Thus, long-term 
historical sunspot data are still an important part of present science 
\citep{Diercke2015, Pavai2016}. Nowadays, space observations monitor the Sun 
continuously with high cadence (tens of seconds) and high-spatial resolution 
(about one second of arc), e.g., the Solar Dynamic Observatory \citep[SDO, 
][]{Pesnell2012} with the onboard instruments Atmospheric Imaging Assambly 
\citep[AIA, ][]{Lemen2012} and Helioseismic and Magnetic Imager \citep[HMI, 
][]{Scherrer2012}. In addition to space instruments, full-disk observations from 
Earth are crucial to understand solar dynamics and activity, e.g., the full-disk 
H$\alpha$~$\lambda$6562.8~\AA\ observations, which are part of the Global 
H$\alpha$ Network \citep{Steinegger2000b}. This network initially combined 
full-disk H$\alpha$ observations from Big Bear Solar Observatory 
\citep[BBSO,][]{Denker1999} in California, Kanzelh\"ohe Solar Observatory 
\citep[KSO,][]{Otruba1999} in Austria, and Yunnan Astronomical 
Observatory~(YNAO) in China. Today, a total of seven facilities around the Earth 
monitor the Sun in H$\alpha$ as part of the network bridging the night gap. 
Furthermore, the Global Oscillation Network Group  \citep[GONG,][]{Harvey1996} 
of the U.S.\ National Solar Observatory (NSO) hosts a similar network. This 
network consists of six facilities across the globe (Spain, U.S.A., Chile, 
Australia, and India) with the aim to study the internal structure of the Sun 
and its dynamics using helioseismology. GONG also provides full-disk H$\alpha$ 
observations and magnetograms.

Along with H$\alpha$ observations, KSO archives include full-disk images of the 
Sun in the continuum and the \mbox{Ca\,\textsc{ii}\,K} line core, as well as 
drawings of sunspots. The Precision Solar Photometric Telescope 
\citep[PSPT,][]{Coulter1996} at Mauna Loa Solar Observatory (MLSO) provides 
full-disk observations in the blue and red continuum, broad-band observations of 
the \mbox{Ca\,\textsc{ii}\,K}~$\lambda$3933.7~\AA\ line, and narrow-band 
observations in the wings of this line. The Full-Disk Patrol 
\citep[FDP,][]{Keller2003} telescope provides full-disk observations at 
high-temporal cadence. Besides H$\alpha$ images, observations in the 
\mbox{Ca\,\textsc{ii}\,K} line, in the 
\mbox{He\,\textsc{i}}~$\lambda$10\,830\,\AA\ triplet, in the continuum, and at 
other photospheric wavelength positions are possible. Doppler velocity maps can 
be created as well using the H$\alpha$ spectral information. The Chromospheric 
Helium Imaging Photometer \citep[CHIP,][]{Elmore1998} at Mauna Loa was operated 
from 1996 until 2013 by the High-Altitude Observatory (HAO). It recorded every 
3\,min full-disk observations of the Sun at seven wavelength positions around 
the \mbox{He\,\textsc{i}}~$\lambda$10\,830\,\AA.

This study concerns another full-disk imager, i.e., the Chromospheric Telescope 
\citep[ChroTel,][]{Kentischer2008, Bethge2011} on Tenerife, Spain, which is 
mounted on the terrace of the Vacuum Tower Telescope 
\citep[VTT,][]{vonderLuehe1998}. It is based on the model of CHIP and also a  
Lyot filter for  \mbox{He\,\textsc{i}} was provided by HAO.  ChroTel is a 
robotic telescope with an aperture of 10~cm and observes at three different 
chromospheric wavelengths: H$\alpha$~$\lambda$6562.8~\AA, 
\mbox{Ca\,\textsc{ii}\,K}~$\lambda$3933.7~\AA, and 
\mbox{He\,\textsc{i}}~$\lambda$10\,830~\AA. In proximity of the infrared 
\mbox{He\,\textsc{i}} line, the instrument is tuned to seven different filter 
positions around $\pm$3~\AA\ centered on the \mbox{He\,\textsc{i}} line core. 
Thus, Doppler maps can be derived based on these filtergrams. The filter 
positions include the close-by photospheric \mbox{Si\,\textsc{i}} line and 
telluric lines.  In the regular observing mode of ChroTel, the telescope records 
every three minutes an image in all three wavelength bands. Higher cadences of 
up to 10~s are possible in H$\alpha$ and \mbox{Ca\,\textsc{ii}\,K}, when 
continuous data acquisition is limited to a single channel. 

Numerous previous studies highlight the importance of the triplet of neutral 
helium \mbox{He\,\textsc{i}}~$\lambda$10\,830\,\AA\ which originates in the high 
chromosphere \citep{Kuckein2012phd}. It consists of a blue and red component, 
where the red component is a blend of two lines. The line is a primary choice to 
analyze the chromospheric magnetic field \citep[see][]{Kuckein2012a} and the 
line-of-sight (LOS) velocities in filaments \citep[see][]{Kuckein2012b}. In both 
studies, the authors presented high-resolution spectral observations obtained 
with the Tenerife Infrared Polarimeter \citep[TIP-II,][]{Collados2007} at the 
VTT. A similar study of high-resolution spectroscopy was carried out with the 
GREGOR Infrared Spectrograph \citep[GRIS,][]{Collados2012} analyzing LOS 
velocities of arch filament systems \citep{GonzalezManrique2018}. On the other 
hand, ChroTel observations were used to to analyze LOS velocities during a 
flare., which was triggered by a splitting sunspot \citep{Louis2014}. Compared 
to high-resolution instruments, ChroTel offers a unique advantage of full-disk 
Dopplergrams in the  \mbox{He\,\textsc{i}} line, which allows us to relate the 
high-resolution observations with the surrounding structures.  Furthermore, 
full-disk  \mbox{He\,\textsc{i}} Dopplergrams enable us to study large-scale 
features such as polar crown filaments over several days continuously. Given the 
wealth of data that ChroTel provides, a method to calibrate the filtergrams and 
produce full-disk Dopplergrams is necessary to fully exploit this data set.

In the following, we present a method to calibrate the ChroTel 
\mbox{He\,\textsc{i}} filtergrams, i.e., the removal of optical aberrations with 
the help of Zernike polynomials. The calibration is necessary because the filter 
transmission introduces a non-uniformity in the intensity distribution of each 
flat-fielded and limb-darkening-corrected filtergram.  In Sect.~\ref{cha:obs}, 
we introduce the ChroTel observations in the \mbox{He\,\textsc{i}} wavelength 
band along with the basic pre-calibration steps. In Sect.~\ref{cha:meth}, we 
describe the method to calibrate the filtergrams yielding a uniform intensity 
background. In addition, we discuss the derived Doppler velocities from the 
ChroTel filtergrams and compare them with corresponding Doppler velocities from 
high-resolution observations of GRIS for a selected sample data set followed by 
a discussion.

  
\section{Observations} \label{cha:obs}

   \begin{figure}[t]
   \centering
   \includegraphics[width=\hsize]{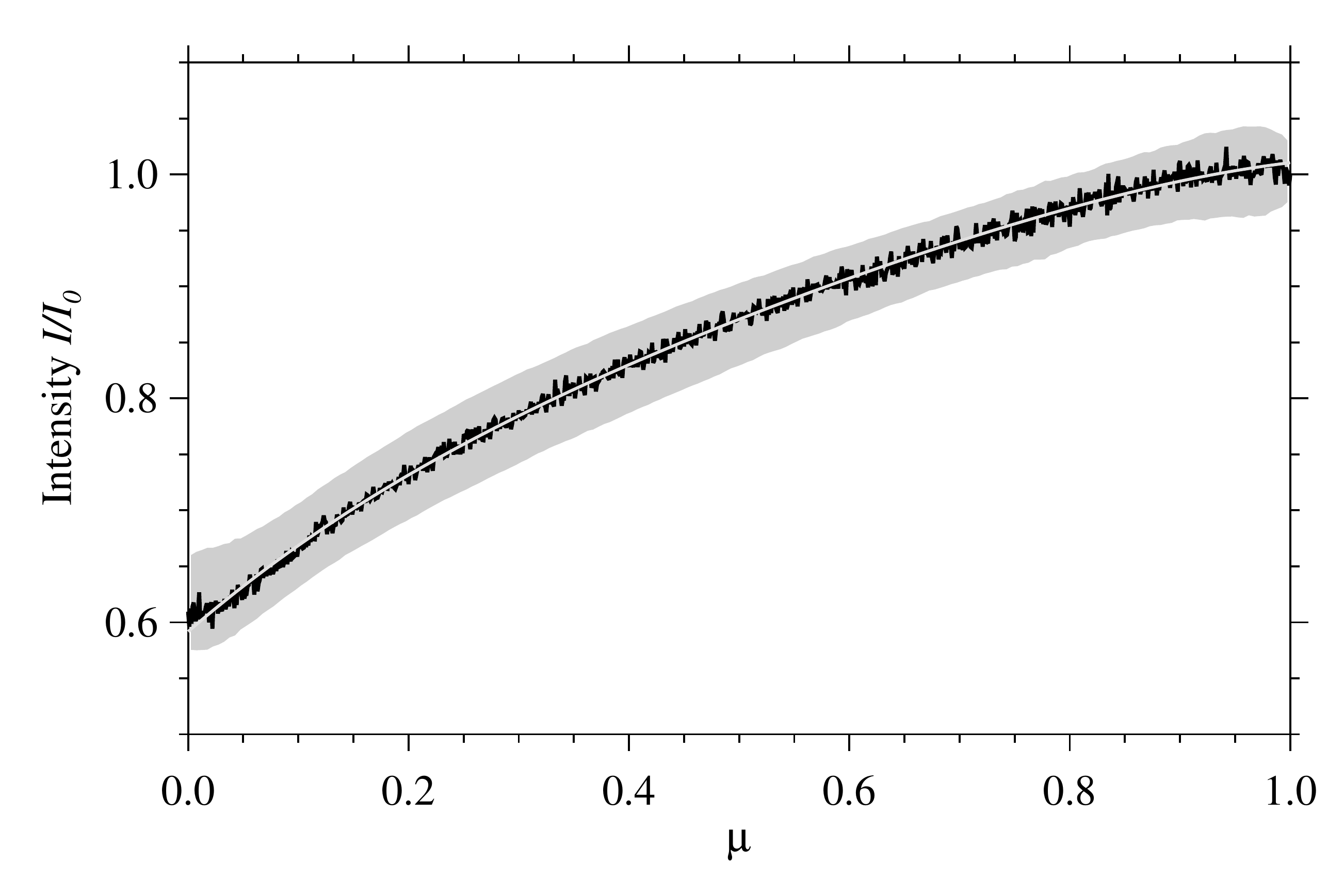}
      \caption{Intensity profile used for the limb-darkening correction of the  
\mbox{He\,\textsc{i}} filtergram at 10:09~UT on 2018 March~23. The intensity $I 
/I_0$, which was normalized to the quiet-Sun intensity at disk center $I_0$, is 
plotted against $\mu = \cos\theta$, where $\theta$ is the heliocentric angle. 
The mean local intensity is plotted in black and the area within three standard 
deviations of the mean local intensity is shaded in gray. A 
$4^\mathrm{th}$-order polynomial fit is overplotted as a white curve.}
         \label{Fig:limb}
   \end{figure}

In the following sections, we introduce the data sets, which we used 
from ChroTel as well as the data set from the high-resolution spectrograph 
GRIS, 
which was used to validate the results derived from the ChroTel observations. 
Furthermore, we discuss shortly the ChroTel image quality of the seven 
filtergrams in \mbox{He\,\textsc{i}} over the day for one data set.

   \begin{figure*}[t]
   \centering
   \includegraphics[width=\textwidth]{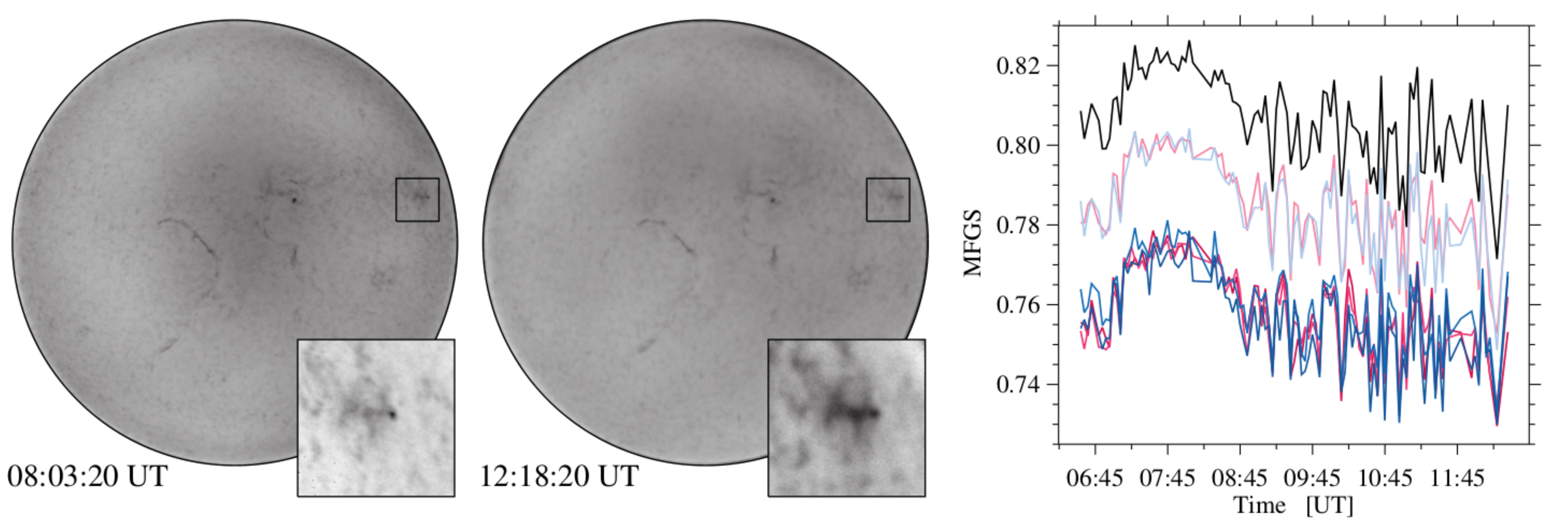}
   \caption{Image quality based on the Median Filter-Gradient Similarity (MFGS) 
method for the time-series on 2017 June~20. The best (\textit{left}) and worst 
(\textit{middle}) image of the time-series were recorded at 08:30:20~UT and 
12:18:20~UT, respectively. Both images are corrected for limb darkening. Active 
region NOAA 12663 is marked with a black square and shown enlarged in the lower 
right corner. Here, the image quality between both images is also visibly 
different. The MFGS values between 06:33\,UT and 12:30\,UT (right) were plotted 
over time for all seven wavelength positions. The black profile 
represents the central filtergram, and blue and red colors correspond to 
filtergrams with shorter and longer wavelengths respectively. Lighter shades of 
blue and red indicate closer proximity to the \mbox{He\,\textsc{i}} line core. 
(different). For clarity, the MFGS curves for the central filtergram 
\#4 is shifted up by an offset of $+0.1$, and the curves for filtergrams \#1, 
\#2, \#6, and \#7 are shifted down by an offset of $-0.1$.}
    \label{Fig:quality}%
    \end{figure*}

\subsection{Observations of the Chromospheric Telescope}

We selected three data sets of ChroTel with different solar activity levels for 
the following investigation. Data representing the active Sun was taken on 
2014~May~12, whereas low solar activity was encountered during quiet-Sun 
conditions on 2018~March~23. The data recorded on 2017~June~20 are a good 
example for typical activity levels. 

ChroTel acquires filtergrams at seven wavelength positions in and around the 
\mbox{He\,\textsc{i}} triplet \citep{Bethge2011}. The filtergrams are numbered 
\#1\,--\,7 in order of increasing wavelength, with \#4 being the central 
filtergram. The central wavelengths for the seven filter position are 
10\,827.45\,\AA, 10\,828.47\,\AA, 10\,829.60\,\AA, 10\,830.30\,\AA, 
10\,831.00\,\AA, 10\,832.13\,\AA, and 10\,833.15\,\AA. Level 1.0 data are 
available online, which are  flat-field corrected and converted to FITS format 
\citep{Wells1981, Hanisch2001}. Each set of filtergrams in level 1.0 data is 
ordered with decreasing wavelength. However, we present all results \#1\,--\,7 
with increasing wavelength for consistency. The 2048$\times$2048-pixel images 
are rotated and rescaled so that the radius corresponds to $r=1000$~pixels, 
which yields an image scale of about 0.96\arcsec~pixel$^{-1}$.  To account for 
limb-darkening, we derive an average intensity profile for each image, as 
described in \citet{Diercke2018}. The intensity values $I$ for each pixel are 
plotted against $\mu = \cos\theta$, where $\theta$ is the heliocentric angle. 
The intensity profile is fitted with a $4^\mathrm{th}$-order polynomial
\begin{eqnarray}
I(\mu)= c_0 + c_1 \cdot \mu + c_2 \cdot \mu^2 + c_3 \cdot \mu^3 +c_4 \cdot 
\mu^4.
\end{eqnarray}
In the next step, the profile is expanded to a 2D map containing the 
limb-darkening values for the whole solar disk. After division of the 
\mbox{He\,\textsc{i}} filtergrams by the 2D limb-darkening function, dark 
sunspots and bright areas are removed by masking these region. The 
$4^\mathrm{th}$-order polynomial fit is then repeated for quiet-Sun regions 
only. An example of the fitted profile is shown in Fig.~{\ref{Fig:limb}\ns}. 
The 2D expansion of this profile is again used for the limb-darkening correction 
of the filtergrams. 

\subsection{Observations of GREGOR Infrared Spectrograph}
On 2017 June~20, active region NOAA~12663 was observed with GRIS, which is 
located close to the west limb (Fig.~\ref{Fig:quality}\ns). The slit length of 
GRIS corresponds to 66.3\arcsec\ with a sampling of 0.136\arcsec~pixel$^{-1}$ 
along the slit. The spectral window covers about 18~\AA, which includes the 
chromospheric \mbox{He\,\textsc{i}} 10\,830~\AA\ line, the 
\mbox{Si\,\textsc{i}}, the \mbox{Ca\,\textsc{i}}, and several telluric lines. 
The dispersion is about 18~m\AA\ pixel$^{-1}$, and the number of spectral points 
is $N_\lambda = 1010$. The spatial scan was taken from 08:33~UT to 09:00~UT, 
covering a field-of-view (FOV) of $20\arcsec\times59\arcsec$. These data will be 
used in Sect.~\ref{cha:dop} to validate the Doppler velocities derived from 
ChroTel filtergrams.

\subsection{Image quality}

The best image in each data set was identified with the Median Filter-Gradient 
Similarity \citep[MFGS,][]{Deng2015} method, which compares the magnitude 
gradient of the original image to that of its median-filtered counterpart. This 
metric is implemented in the sTools software package \citep{Kuckein2017IAU, 
Denker2018}. The evolution of MFGS values with time over the course of an 
observing day is shown as an example in Fig.~{\ref{Fig:quality}\ns} for the 
data set obtained on 2017 June~20. The best seeing conditions and thus the best 
images are encountered in the morning for all seven wavelength positions. The 
central filtergram has a maximum MFGS value of about 0.82 at 08:03:20~UT. The 
image is visibly sharper compared to the worst central filtergram at 12:18:20~UT 
with an MFGS value of only 0.76, which appears blurred and unsharp. Increasing 
turbulence, when the ground is heated by the Sun, and decreasing air mass are 
competing factors for the prevailing seeing conditions. Thus, mountain-island 
observatory sites such as Observatorio del Teide experience the best and most 
stable seeing conditions in the early morning hours. This applies to both 
synoptic full-disk observations with ChroTel and high-resolution observations 
with the VTT and the GREGOR solar telescope \citep{Schmidt2012}. Comparing MFGS 
values of filtergrams at different wavelength positions, the central filtergram 
reaches highest values (see black profile in the right panel of 
Fig.~{\ref{Fig:quality}\ns}). The blue and red line-wing filtergrams (light 
blue and red profiles) exhibit the next highest MFGS values. The lowest values 
are encountered in the far line-wing and continuum filtergrams. The drop of 
MFGS values with distance from the \mbox{He\,\textsc{i}} line core can be 
explained by the diminishing fine-structure contents of the filtergrams, i.e., 
the MFGS method is structure dependent, which has to be considered when 
comparing different types of images. Nonetheless, all MFGS time profiles are 
tightly correlated with correlation values between 0.89 and 0.95 and display the 
same trend, i.e., improving seeing conditions until about 08:00~UT, followed by 
a decrease until 09:00~UT. This time interval clearly demonstrates the changing 
contributions of air mass and (ground-layer) turbulence to the prevailing seeing 
conditions. After 09:00~UT, larger MFGS variations on shorter time-scales are 
observed, which are likely linked to local seeing in the vicinity of ChroTel's 
roof-mounted turret.


   \begin{figure*}[t]
   \centering
   \includegraphics[width=\textwidth]{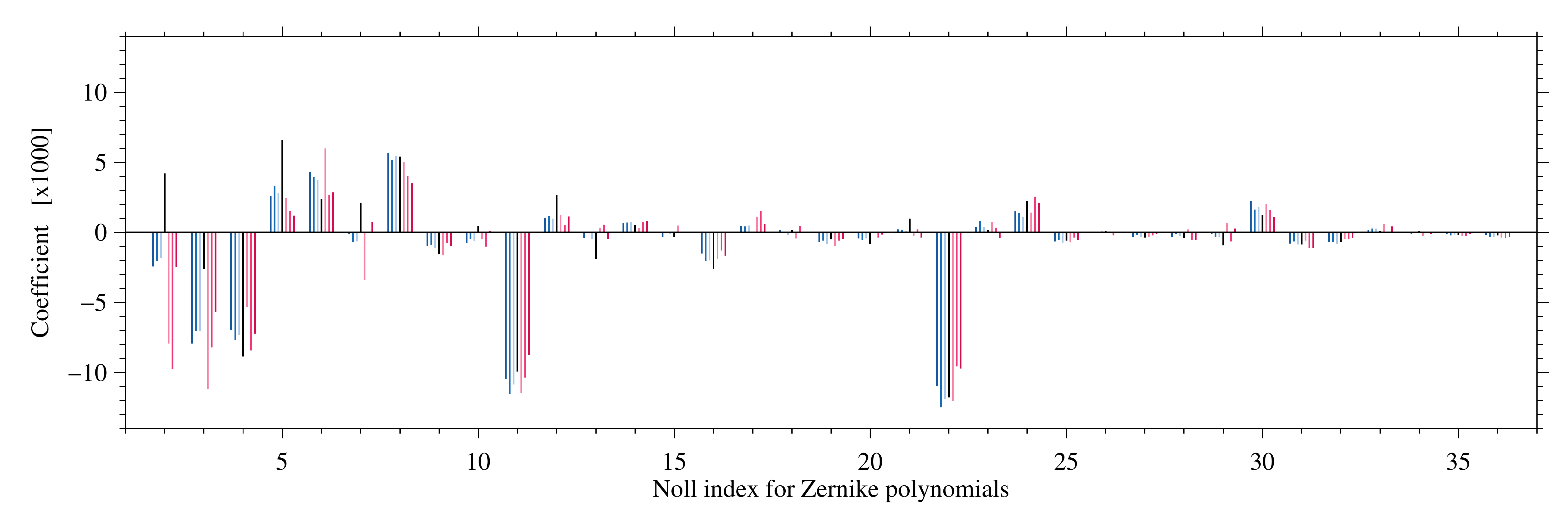}
   \caption{Coefficients of 36 Zernike polynomials fitted to 
\mbox{He\,\textsc{i}} filtergrams at seven wavelength positions for the best 
quiet-Sun data set on 2018~March~23. For each position, the coefficients are 
displayed in increasing wavelength from left to right. Black refers to the 
line-core filtergram, while blue and red correspond to the red and blue line 
wings (see color-coding in Fig.~{\ref{Fig:quality}\ns}).}
    \label{cf_plot}%
    \end{figure*}

   \begin{figure*}[t]
   \centering
   \includegraphics[width=\textwidth]{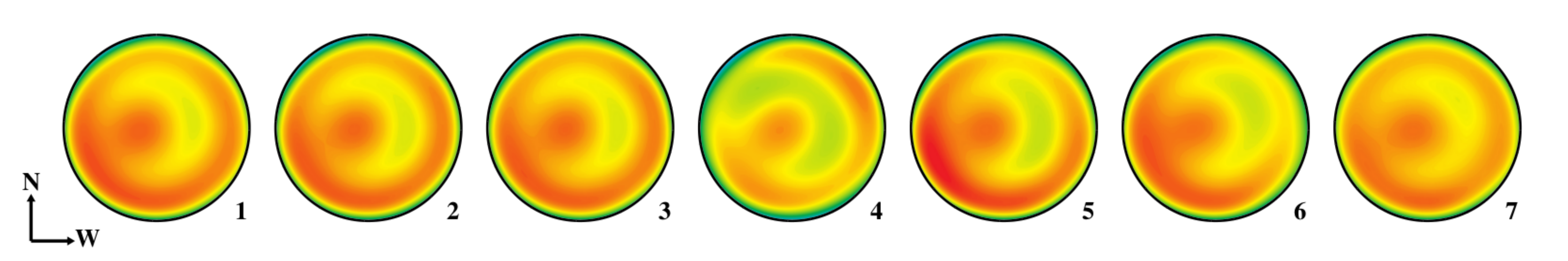}
   \caption{2D reconstruction of the background based on the Zernike 
coefficients depicted in Fig.~{\ref{cf_plot}\ns} for the best quiet-Sun data set 
on 2018~March~23. Each filtergram is labeled \#1\,--\,7 according to increasing 
wavelength. The images are scaled in the same interval $[0.75,\,1.07]$.}
    \label{zkfit_plot}%
    \end{figure*}

\section{Methods} \label{cha:meth}

\subsection{Zernike polynomials}

The limb-darkening corrected filtergrams in Fig.~{\ref{Fig:quality}\ns} exhibit 
an uneven background, which is likely introduced by time-dependent filter 
transmission and image rotation caused by the alt-azimuthal mount of the turret 
system. The central part of the image is darker with a brighter ring surrounding 
it. The uneven background is visible in each filtergram and persists throughout 
the time-series but continuously changes its appearance. These intensity changes 
across the solar disk and from filtergram to filtergram hamper any efforts to 
reliably calculate Doppler velocities. 

To remedy this shortcoming, we developed a calibration process for ChroTel 
\mbox{He\,\textsc{i}} filtergrams using Zernike polynomials to characterize this 
uneven background. Zernike polynomials are a sequence of orthogonal polynomials 
on the unit disk, which can be expressed in polar coordinates as a product of 
angular and radial functions. They are widely used to describe aberrations of 
optical systems. Since they are orthogonal, Zernike polynomials are often chosen 
as a basis to represent properties of a circular image with no redundancy. In 
the following, we will show that Zernike polynomials are well-suited to remove 
artifacts from solar full-disk images -- with applications well beyond the 
ChroTel data at hand. 

\citet{Noll1976} modified the polynomials with a special normalization scheme 
and introduced a new index system. The mode-ordering number $j$ is a function of 
the radial degree $n$ and azimuthal frequency $m$. The filter transmission is 
expressed in terms of Zernike polynomials up to mode-ordering number $j=36$, 
which corresponds to radial degrees $n \le 8$ and azimuthal frequencies $|m| \le 
8$.

   \begin{figure*}[t]
   \centering
   \includegraphics[width=\textwidth]{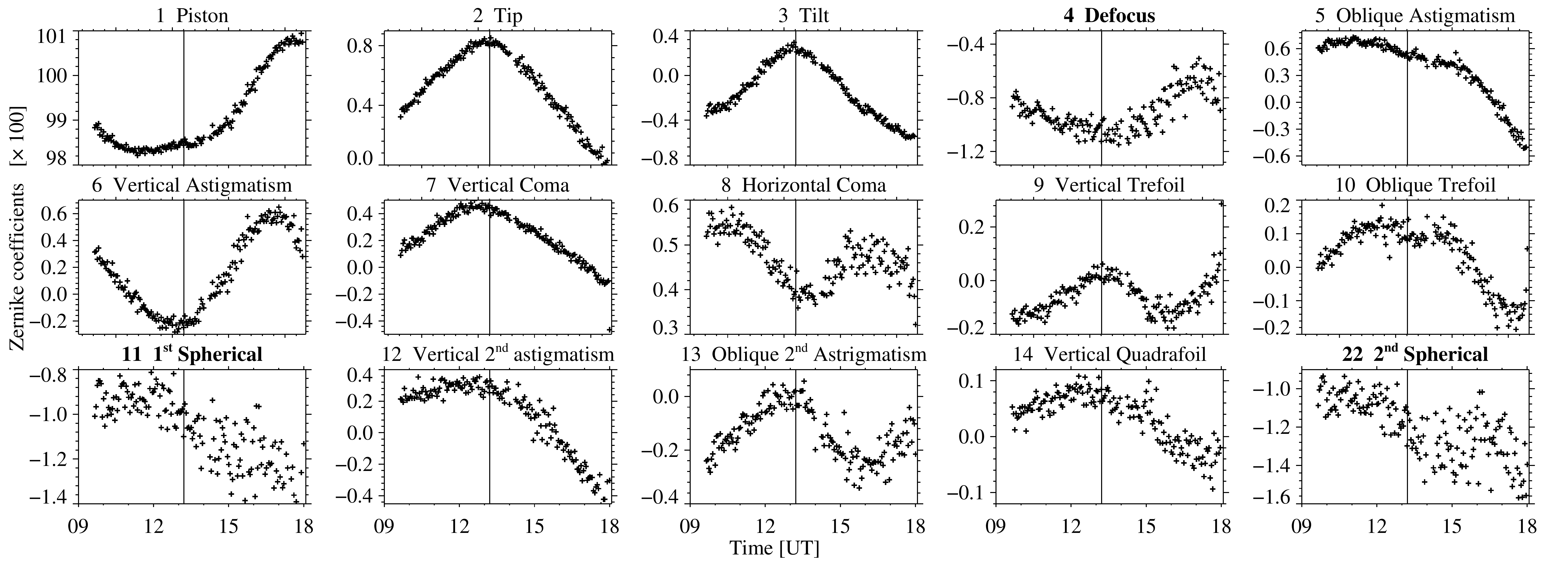}
   \caption{Temporal evolution of Zernike coefficients corresponding to the 
\mbox{He\,\textsc{i}} line-core filtergram. The images were taken from 09:39~UT 
to 17:57~UT on 2018 March~23. Noon local time at the Observatorio del Teide on 
Tenerife is marked with a vertical line. The title on each panel includes the 
Noll index $j$ followed by the names of the corresponding optical aberrations. 
The radial polynomials $j=4$, 11, and 22 are marked in bold.}
    \label{cf_time_plot}%
    \end{figure*}

   \begin{SCfigure*}
   \centering
   \includegraphics[width=0.67\textwidth]{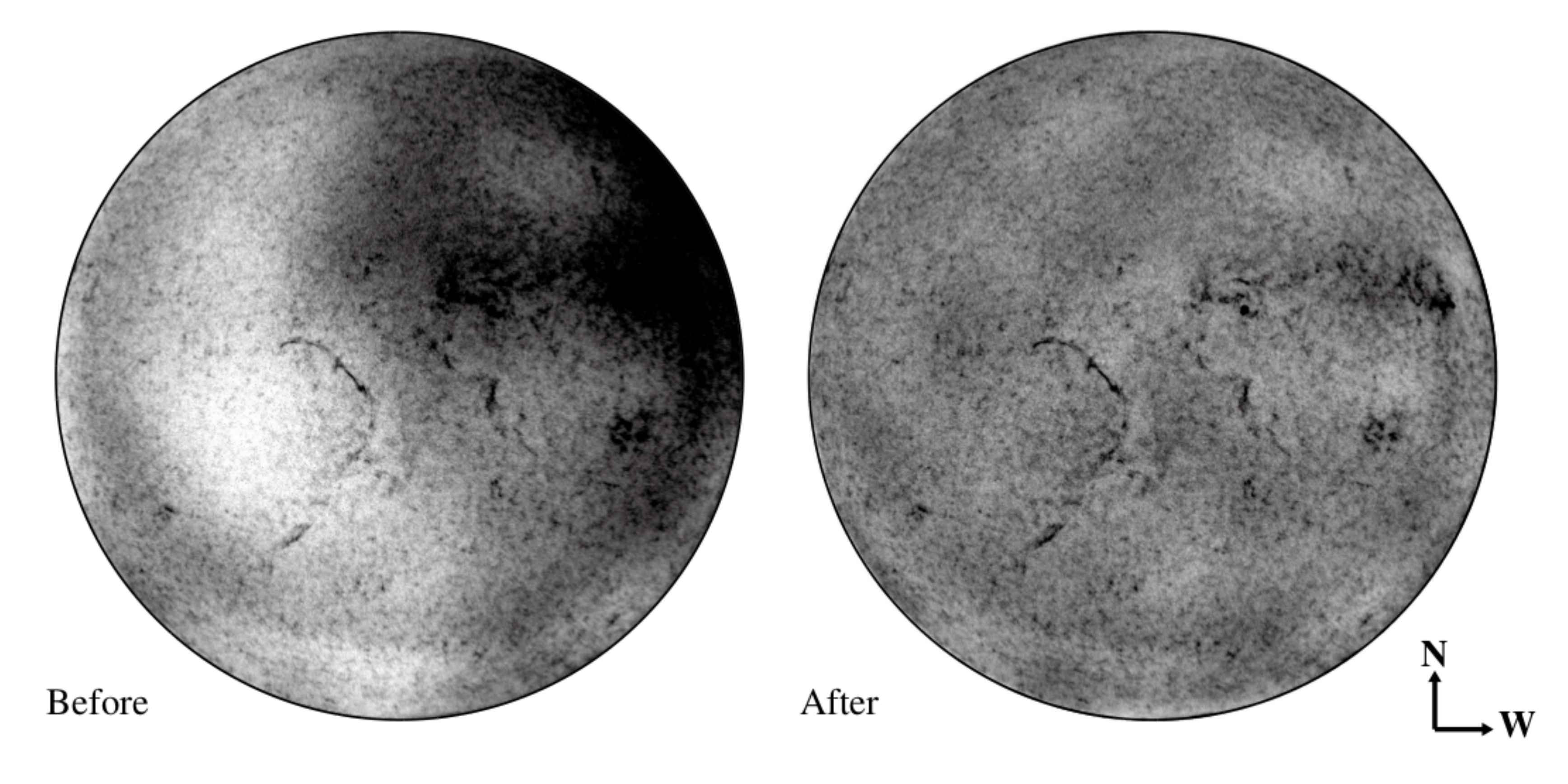}
      \caption{Comparison of a ChroTel filtergram before (\textit{left}) and 
after (\textit{right}) calibration using the Zernike reconstruction of the 
background. The \mbox{He\,\textsc{i}} filtergram was taken in the red line wing 
(\#5) at 08:03:20~UT on 2017~June~20 at a time of average solar activity.} 
         \label{before_after_plot}
   \end{SCfigure*}


\subsection{Calibration process}\label{cha:cal}

Since the solar radius in ChroTel filtergrams is rescaled to 1000 pixels, we 
inscribe the circular Zernike polynomials in a square image with $2000 \times 
2000$~pixels. Zernike polynomials are generated with routines written in the 
Interactive Data Language (IDL) included in the sTools software repository, 
which were originally developed by \citet{Denker2005} for measuring the quality 
of Fabry-P\'erot etalons. Fitting arbitrary intensity distributions across a 
disk, like those of ChroTel \mbox{He\,\textsc{i}} filtergrams, is also 
implemented in these IDL routines. Since the geometry of Zernike polynomials 
matches that of the solar disk in \mbox{He\,\textsc{i}} filtergrams, the 
pixel-to-pixel correspondence can be exploited. The intensity values of the 
limb-darkening-corrected filtergrams are placed in a column vector $b$ with 
about $3.14 \times 10^6$ elements. Similarly, the first 36 Zernike polynomials 
form the 36 columns, with $3.14 \times 10^6$ elements each, of the matrix $A$. 
The intensity values in the solar disk are then fitted with a linear combination 
of 36 Zernike polynomials, which becomes a linear least-squares problem $A \cdot 
x = b$ using matrix-vector multiplication. Finally, singular value decomposition 
\citep[SVD,][]{Press2002} and back-substitution is used to obtain the 36 
coefficients $x$ of the Zernike polynomial fit. This calibration has one caveat 
because it applies only to quiet-Sun regions, where \mbox{He\,\textsc{i}} 
absorption profiles are not present. Consequently, active regions and filaments 
with strong \mbox{He\,\textsc{i}} absorption features have to be excluded from 
the fitting procedure. This can be accomplished by masking these regions in 
\mbox{He\,\textsc{i}} line-core filtergrams using intensity thresholds and 
morphological image processing. This process is carried out in three steps. 
First, the background is computed, including the active regions and filaments, 
and removed from the line-core filtergram. Second, the corrected filtergram is 
used to determine the mask for active regions and filtergrams. Third, the 
background is computed excluding active regions and filaments. The mask 
determined in the second step is applied to all filtergrams at the seven 
wavelength positions.  

To determine the optimal number of Zernike polynomials to be used in the fit, we 
tested the calibration method on the best quiet-Sun filtergrams which was  taken 
at 10:09~UT on 2018 March~23.  On this day, no active regions or filaments were 
present and thus all pixels in the solar disk were used in the fitting 
procedure. Figure~{\ref{cf_plot}\ns} displays the coefficients for each Zernike 
polynomial for the set of seven filtergrams. The piston term $j=1$, representing 
the average intensity, is not shown. The strongest modes are $j=4$, 11, and 22. 
These Zernike polynomials have only a radial component. In optical terminology, 
$j=4$ refers to the defocus term, $j=11$ represents the primary and $j=22$ the 
secondary spherical aberration \citep{Noll1976}. Most sets of the seven Zernike 
coefficients are similar but not identical for each Noll index $j$. This points 
to a common origin for the observed variation of the filter transmission. The 
variation is sufficient in magnitude to necessitate calibrating the filtergrams 
individually for each wavelength position. Figure~{\ref{zkfit_plot}\ns} 
displays the 2D reconstruction of the background using the Zernike coefficients 
plotted in {Fig.~\ref{cf_plot}\ns}. 

   \begin{figure*}[t]
   \centering
   \includegraphics[width=\textwidth]{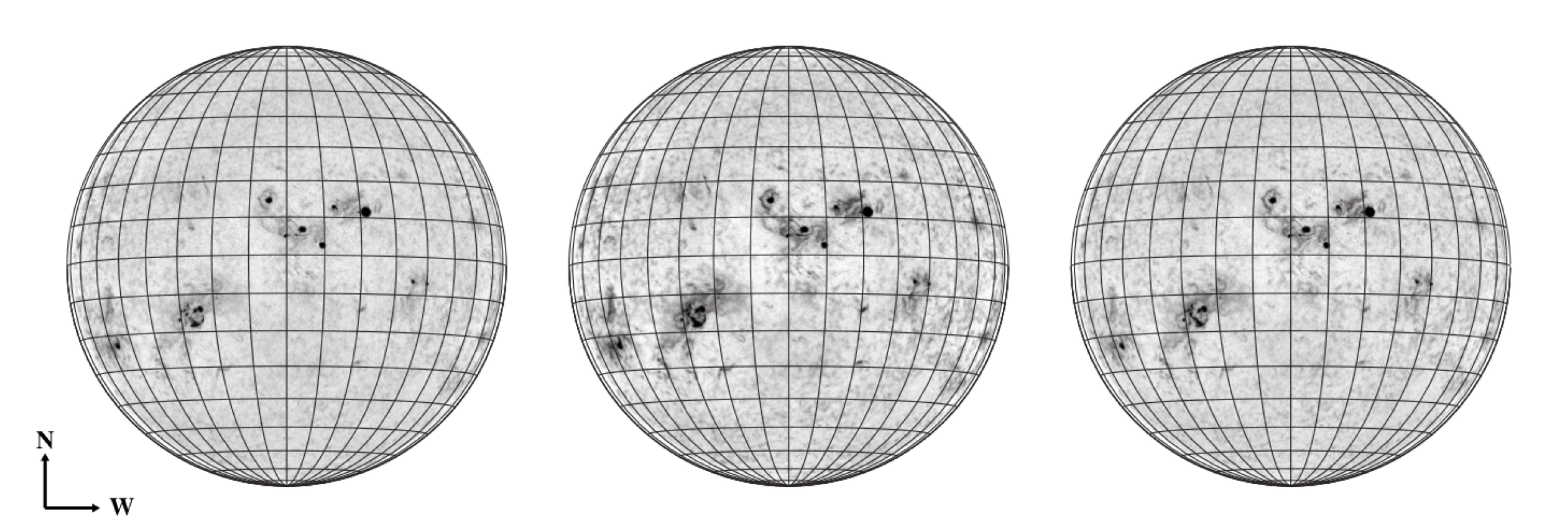}
   \caption{Calibrated ChroTel \mbox{He\,\textsc{i}} filtergrams taken on 2014 
May~12: blue line-wing \#3, line-core \#4, and red line-wing \#5  filtergram 
(\textit{from left to right}), respectively. The image intensity was scaled in 
the interval [0.86, 1.03] and a Stonyhurst grid was superposed on the solar 
disk.}
    \label{active_3p_plot}%
    \end{figure*}

The Zernike reconstructions of the background identify a dark circle near the 
center and an asymmetrical ring-shaped structure on the disk, both of which 
correspond to strong radial modes. Attempts to fit more Zernike polynomials 
reveal higher-order radial modes, while the non-radial polynomials become 
increasingly smaller. These higher-order polynomials characterize small-scale 
structures near the edge of the disk or high-frequency radial variations. 
However, only broad trends across the whole disk are subject of the calibration 
process. Thus, the limit is appropriately set at $j=36$, which is also supported 
by visual inspection of the corrected filtergrams. The orthogonality of Zernike 
polynomials ensures that the first 36 coefficients will not be affected if 
higher order terms are neglected. The Zernike fits identify different structures 
in each filtergram, even though similarities prevail on larger scales. 
Filtergram \#4 refers to the core of the \mbox{He\,\textsc{i}} line, so that the 
disk has a lower average intensity than the other filtergrams. The filtergrams 
were taken with tunable Lyot-type filters, which means that the wavelength 
position of the transmission peak is shifted very rapidly by applying a variable 
voltage to liquid crystal retarders between the  birefringent  stages 
\citep{Bethge2010phd}. This may introduce additional changes between 
filtergrams, which become visible in the Zernike reconstructions of the 
background.  

In addition, we investigated the evolution of the Zernike coefficients, again 
using the complete quiet-Sun data set on 2018 March~23. 
Figure~{\ref{cf_time_plot}\ns} shows the time-dependent variation of the first 
14 Zernike coefficients, as well as the coefficients of the radial mode $j = 
22$. We only display the coefficients for the time-series of the 
\mbox{He\,\textsc{i}} line-core filtergrams because they are representative for 
the other filter positions as well. The three radial modes $j=4$, 11, and 22 
show a large scatter, while the non-radial coefficients exhibit a clear trend as 
the day progressed. The piston term $j=1$ characterizes the average intensity of 
the solar disk, which is close to unity after limb-darkening correction. Most 
coefficients, e.g. $j=2$, 3, and 7, peak at local noon. This suggests that 
either changes in the light level or elevation of the Sun are responsible for 
the observed time dependence. ChroTel has an alt-azimuth mount that introduces
image rotation, which is however corrected as part of the data reduction. The
Lyot filter is mounted in the collimated beam, which results in a stationary
pupil. More importantly, the two mirrors of the turret introduce linear
polarization, which  changes over the day. This influences strongly the
transmission of the Lyot filter. A time-lapse movie covering the whole observing
period reveals that the central dark spot (Fig.~{\ref{Fig:quality}\ns}) moves
across the solar disk. The temporal evolution of the coefficients quantitatively
describes how transmission artifacts evolve over time. In principle, the time
dependence of the Zernike coefficients can be modeled since the shape of the
profiles are rather simple. However, analyzing a large sample of daily
\mbox{He\,\textsc{i}} time-series is beyond the scope of this study. Applying the
corrections to a single filtergram takes only about 180\,s on a modern desktop
computer so that processing time-series data is feasible.

To validate the background calibration procedure employing Zernike polynomials, 
we evaluate the statistical moments of the intensity distributions on the solar 
disk. Since strong \mbox{He\,\textsc{i}} absorption features are absent in  
quiet-Sun filtergrams, the intensity values are expected to have a symmetrical 
distribution. Skewness, i.e., a measure of the asymmetry of the intenisty 
distribution, is mainly introduced by variations in filter transmission. We 
compare the skewness of the best quiet-Sun filtergrams from 2018~March~23  
before and after calibration. The calibration reduces the skewness in most 
filtergrams. Relevant for determining Doppler velocities (see 
Sect.~\ref{cha:dop}) are the blue and red line-wing filtergrams, where the 
skewness is reduced from $+0.17$ to $+0.12$ and from $-0.10$ to $-0.02$, 
respectively. The variance of the intensity distribution is also reduced for all 
seven filtergrams. The average variance is $2.1 \times 10^{-4}$ before and $0.67 
\times 10^{-4}$ after calibration. The intensity distributions of the calibrated 
filtergrams are visually more symmetrical and the tails of the distribution are 
more confined. This raises our confidence that large-scale variations introduced 
by the filter transmission were significantly reduced in the filtergrams, 
resulting in a flat background across the solar disk. 

Displaying the calibrated quiet-Sun filtergrams is not very instructive because 
they are void of any significant absorption structures such as active regions 
and filaments. Therefore, the best \mbox{He\,\textsc{i}} red line-wing 
filtergram on 2017~June~20 was chosen for the direct comparison. 
Figure~{\ref{before_after_plot}\ns} clearly demonstrates the improvements after 
limb-darkening correction and calibration of the background. The intensity was 
clipped tightly to display the background variations more prominently. The 
calibration process removed the gradient towards the upper left corner of the 
image and results in an image with even background. The slightly brighter areas 
in the center of the solar disk surrounding the active regions are real and were 
observed in H$\alpha$ line-wing filtergrams as well \citep{Marquette1999}. 
Finally, the calibrated filtergrams \#3\,--\,5 on 2014~May~12 in 
Fig.~{\ref{active_3p_plot}\ns} are displayed in a broader intensity range to 
bring forth details within the active regions. Properly calibrated line-wing 
and line-core filtergrams are the basis for the derivation of Doppler velocities 
in the following section.

   \begin{figure}[t]
   \centering
   \includegraphics[width=\columnwidth]{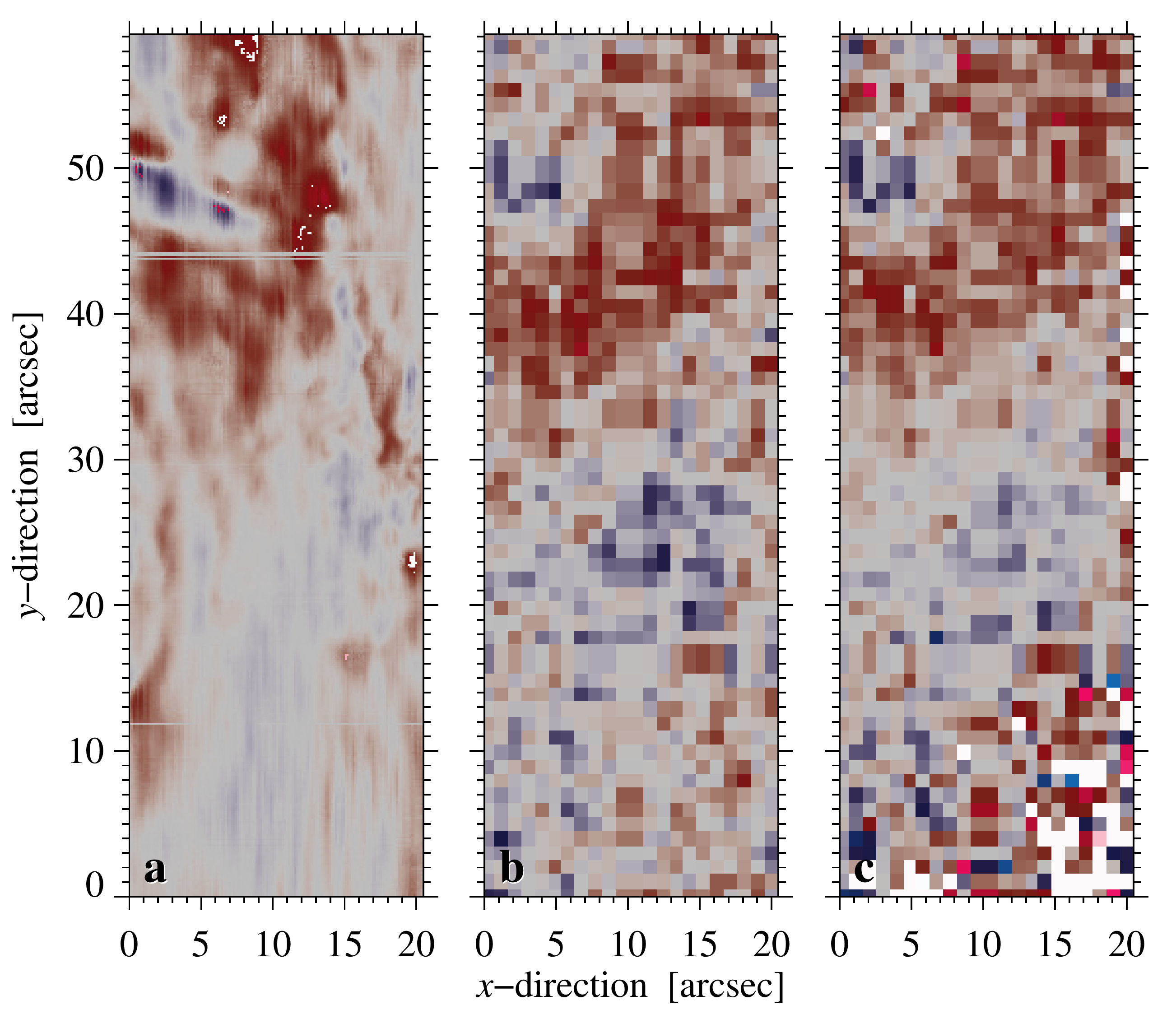}
      \caption{Doppler velocity maps obtained with (a) GRIS and ChroTel, which 
were derived with (b) the difference method and (c) a parabolic fit to the 
central three filtergrams. The difference map is scaled between $\pm$0.1 in 
terms of the normalized quiet-Sun intensity, whereas the other two maps are 
scaled between $\pm$20~km~s$^{-1}$. The ChroTel maps appear pixelated due to the 
considerable difference in image scale.}
      \label{Fig:dop3}
   \end{figure}
%

\subsection{Doppler Velocity} \label{cha:dop}

\citet{Bethge2011} produced Doppler maps from ChroTel filtergrams using a 
center-of-gravity (COG) method to identify the line center. Their COG formula 
contains arbitrary coefficients to account for filter transmission, blending of 
different components of the \mbox{He\,\textsc{i}} line, and other factors that 
affect real data. These coefficients are optimized for each data set by 
comparing ChroTel Doppler maps to co-spatial, high-resolution observations of 
the TIP-II instrument at the VTT. They concluded that for velocities up to about 
20~km~s$^{-1}$ the inner three filtergrams are sufficient, but the coefficients 
are not universally applicable. New data sets may require re-calibration with 
near-infrared spectrographic scans. Because of the scarcity of near-infrared 
spectrographic data, the synoptic observations from ChroTel cannot be fully 
exploited. Furthermore, the time dependence of the Zernike coefficients (see 
Sect.~\ref{cha:cal} and Fig.~{\ref{cf_time_plot}\ns}) implies that such a 
cross-calibration is only valid for a short period of time (tens of minutes).

The calibration of the \mbox{He\,\textsc{i}} filtergrams with the help of 
Zernike polynomials allows us to test two methods to derive Doppler velocities. 
The first method is a simple difference map, where the red line-wing filtergram 
is subtracted from the blue line-wing filtergram. The second method is a 
parabolic fit to the ``line profile'' of the inner three filtergrams. The 
reference Doppler velocities are obtained with GRIS from a high-resolution 
spectrograph scan taken on 2017 June~20 between 08:33~UT and 09:00~UT. The scan 
covers parts of active region NOAA~12663 near the west limb. The Doppler shift 
map from GRIS is compiled by fitting the \mbox{He\,\textsc{i}} line core in each 
pixel with a parabola and determining its minimum position. The corresponding 
Doppler velocity map is shown in Fig.~{\ref{Fig:dop3}\ns}a, scaled between 
$\pm$20~km~s$^{-1}$. 

   \begin{figure}[t]
   \centering
   \includegraphics[width=\columnwidth]{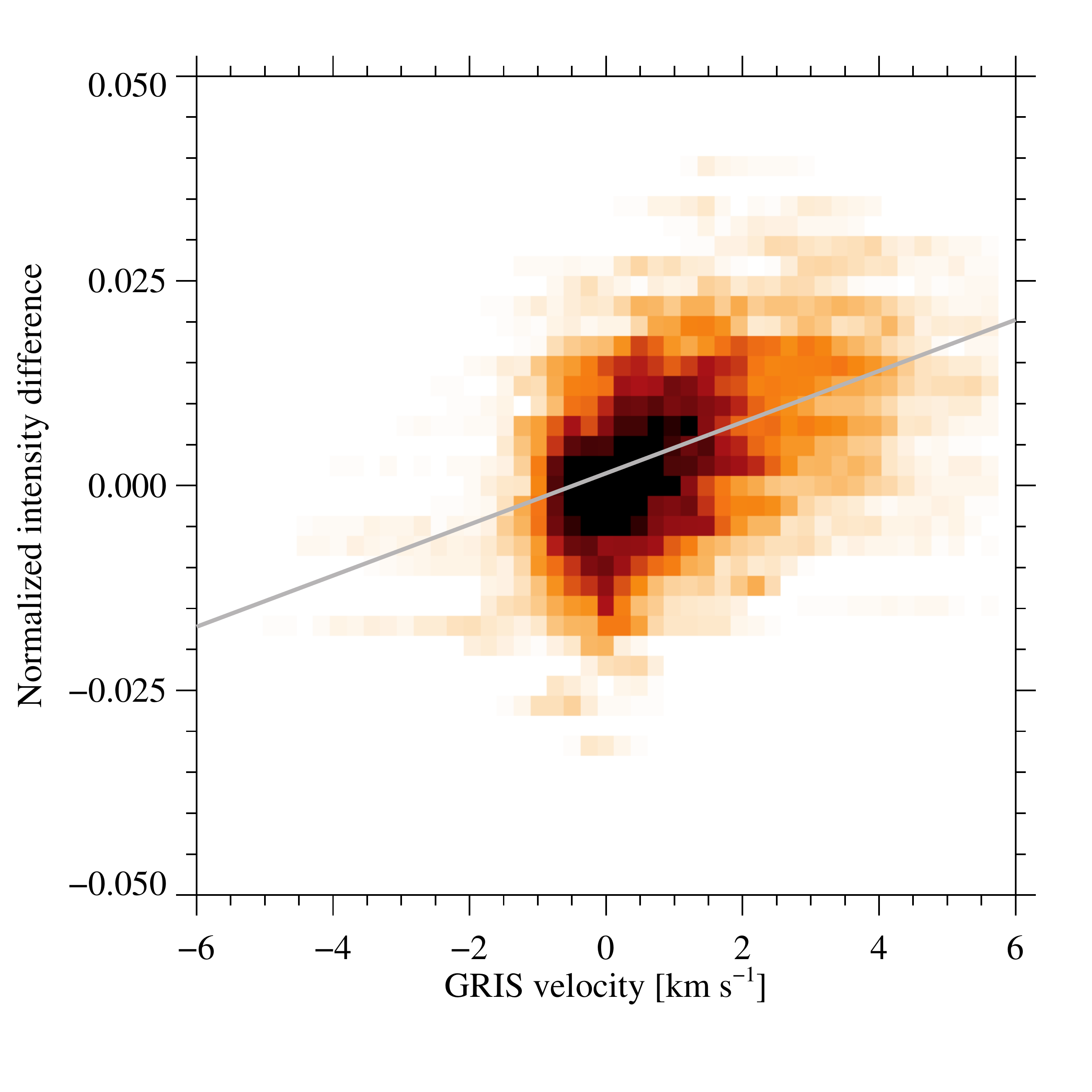}
      \caption{Reference velocity (GRIS) vs.\ velocity analogue $v^\prime$ 
(ChroTel). The light grey line depicts a linear regression with a slope of $m = 
0.0031$\,s~km$^{-1}$ and a vertical intercept of 0.0015. The background image is 
a color-coded 2D histogram.} 
      \label{Fig:gris_vch}
   \end{figure}

   \begin{figure}[t]
   \centering
   \includegraphics[width=\columnwidth]{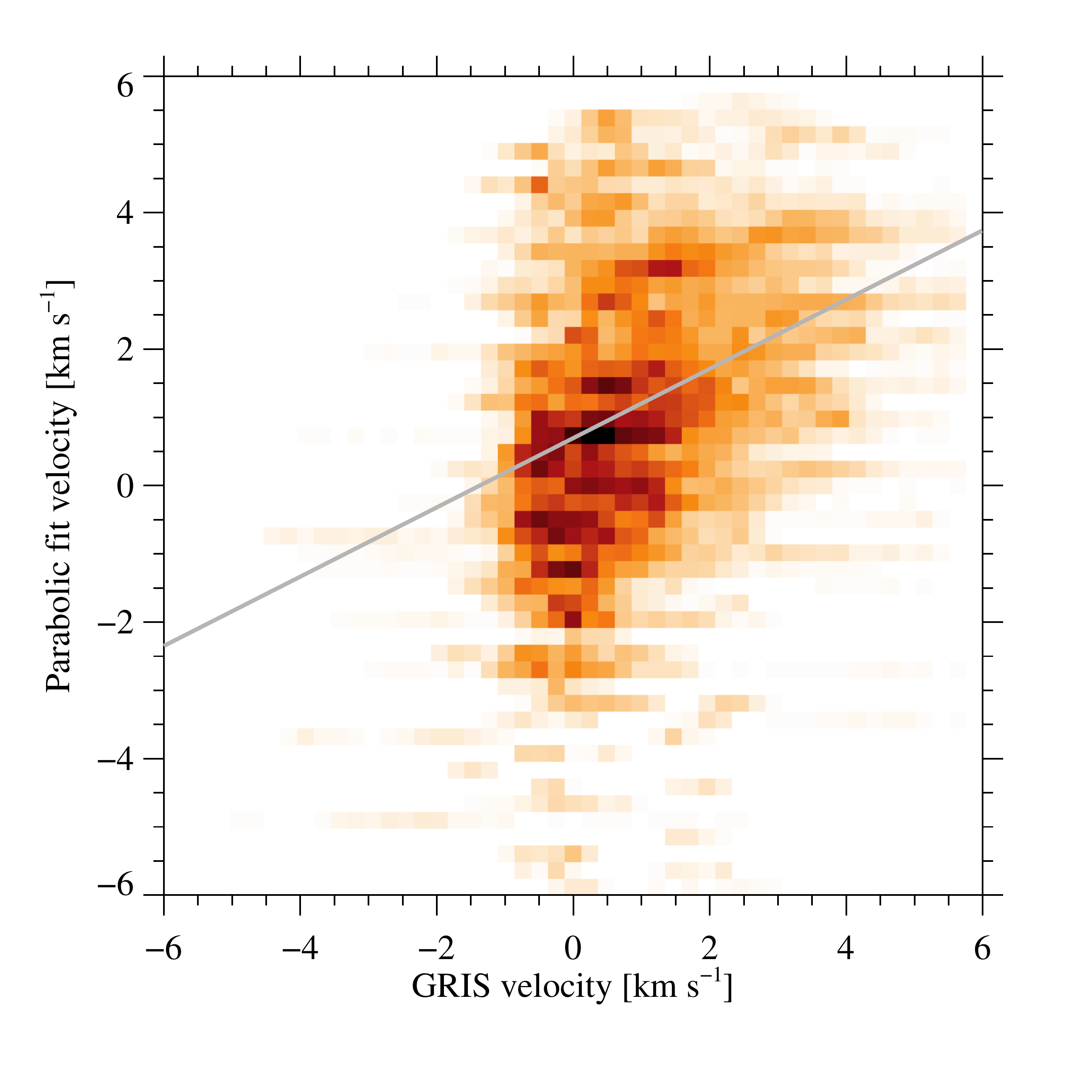}
      \caption{Reference velocity (GRIS) vs.\ velocity derived with a parabolic 
fit (ChroTel) in the velocity range of $\pm$6~km~s$^{-1}$. The light grey line 
depicts a linear regression with a slope of 0.51 and a vertical intercept of 
0.70~km~s$^{-1}$. The background image is a color-coded 2D histogram.}
      \label{Fig:gris_vlos}
   \end{figure}

Multiple ChroTel filtergrams were taken during the scanning period of GRIS, and 
we chose the filtergram with the best seeing conditions at 08:33~UT for 
comparison. The GRIS scanning time is around half an hour, shorter than the time 
scale of the temporal evolution of the Zernike coefficients of about 
2--3\,hours. The region scanned with GRIS is extracted from the ChroTel 
filtergrams and carefully aligned to the GRIS map. In addition to the continuum 
image, the Doppler maps from ChroTel and GRIS are also aligned with respect to 
each other. Figure~{\ref{Fig:dop3}\ns} compares the GRIS Doppler map to ChroTel 
velocities determined with the two different methods.

After the calibration, the quiet-Sun intensity in areas with no 
\mbox{He\,\textsc{i}} absorption is close to unity. The intensity difference 
between the blue line-wing filtergram \#3 and the red line-wing filtergram \#5 
is directly related to the line shift because of the normalization across the 
filtergrams. This normalized intensity difference $\delta I = I_r -I_b$ is 
proportional to the Doppler velocity $v^\prime = C \cdot \delta I$, which is 
similar as described in the method of \citet{Padinhatteeri2010}. The factor $C$ 
will be derived from the comparison with the GRIS Doppler velocities and can 
then be extended to the full-disk data set. The Dopplergrams from the Michalson 
Doppler Imager \citep[MDI,][]{Scherrer1995} on board the Solar and Heliospheric 
Observatory \citep[SoHO,][]{Domingo1995} are produced similarly with a modified 
difference method using four filtergrams in the wing.

\begin{figure}[t]
   \centering
   \includegraphics[width=\columnwidth]{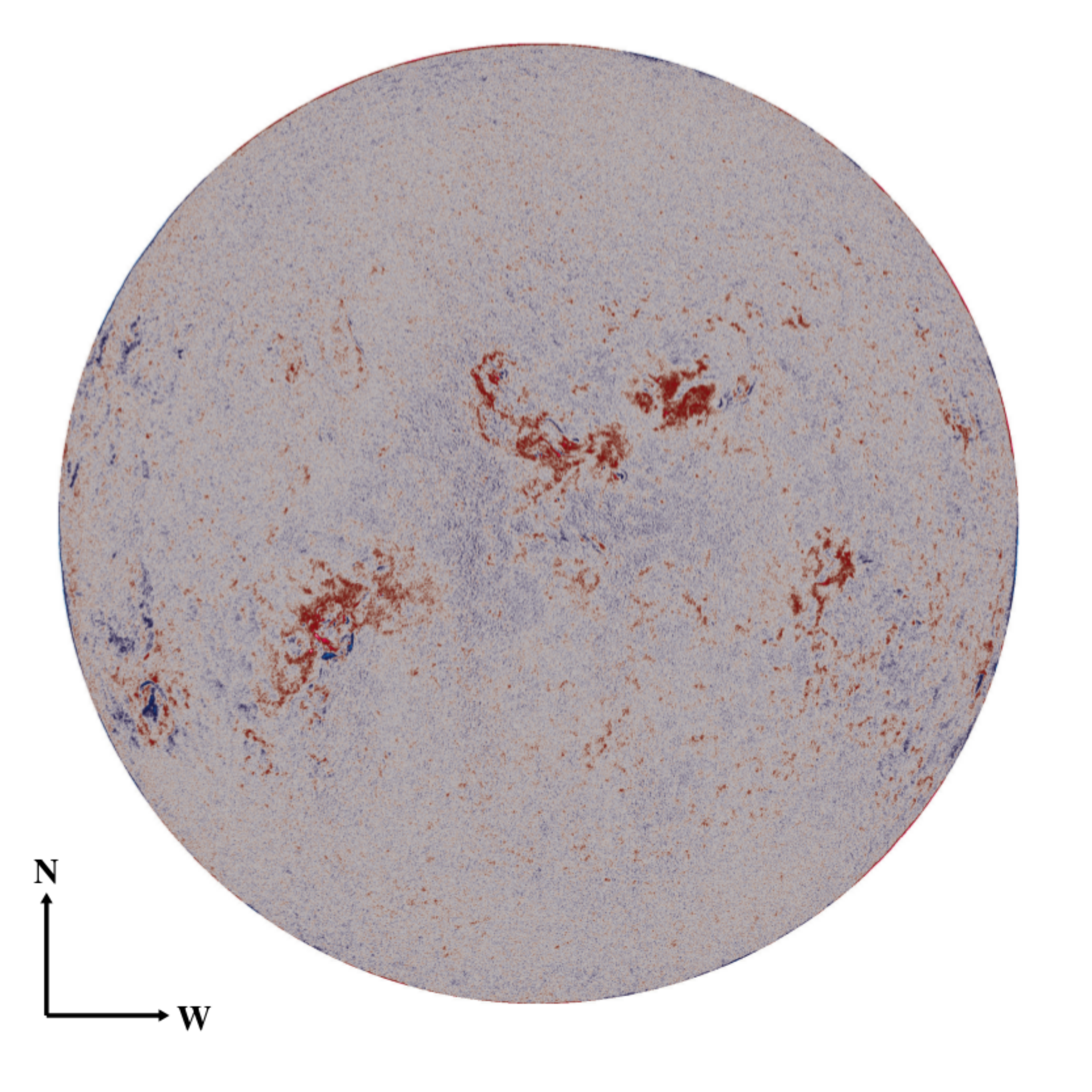}
   \caption{ChroTel full-disk difference maps (red minus blue line-wing 
filtergram) for the best \mbox{He\,\textsc{i}} data set on 2014 May~12. The 
effects of solar differential rotation were removed in the Zernike fitting 
process. The normalized intensity difference was multiplied by the 
factor $C = 320$\,km~s$^{-1}$.The images are scaled between 
$\pm$35\,km\,s$^{-1}$. The  median velocities is about 
$\pm1.6$\,km\,s$^{-1}$ and the velocity outside the active regions is between 
$\pm15$\,km\,s$^{-1}$.}
    \label{dop_disk_3p_plot}%
    \end{figure}

Since the intensity difference is a proxy for line shift, the resulting 
velocities should be proportional to the reference LOS velocity. This difference 
map is presented in Fig.~{\ref{Fig:dop3}\ns}b, scaled between $\pm$0.1. This 
normalized intensity difference has the advantage that it is defined across the 
entire solar disk. In regions without \mbox{He\,\textsc{i}} absorption, the 
intensity of both line-wing filtergrams is normalized to unity, and any 
background artifacts have been removed by the Zernike fitting. Thus, the 
difference is close to zero as is the Doppler velocity. This is confirmed in the 
full-disk difference map in Fig.~{\ref{dop_disk_3p_plot}\ns}. On the other 
hand,  the line shift in regions with \mbox{He\,\textsc{i}} absorption is due to 
both differential rotation and the local plasma motion.

For regions with \mbox{He\,\textsc{i}} absorption, the contribution from 
differential rotation results in a broad trend across the solar disk. The 
Zernike polynomials can identify and remove this from the final calibrated 
filtergram. The caveat is that large active regions with strong absorption are 
excluded from the Zernike polynomial fit. In addition, co-temporal GRIS scans 
cover an area too small to determine differential rotation effects. We need to 
compare the obtained Doppler velocities to other full-disk \mbox{He\,\textsc{i}} 
Doppler maps in order to identify whether differential rotation has been 
removed.

For the second method, the intensity trace of the inner three filtergrams can be 
treated as a low-resolution line profile for each pixel location. For pixels 
with \mbox{He\,\textsc{i}} absorption as determined from the line-core 
filtergram, a parabolic fit allows us to compute more accurately the line center 
and thus the Doppler velocity. We have to exclude some pixels with a negative 
coefficient for the quadratic term, i.e., where the vertex of the parabola is on 
the top. The resulting map is shown in Fig.~{\ref{Fig:dop3}\ns}c, where the 
excluded pixels are displayed in white. The map is scaled between 
$\pm$20~km~s$^{-1}$. 

   \begin{figure*}[t]
   \centering
   \includegraphics[width=\textwidth]{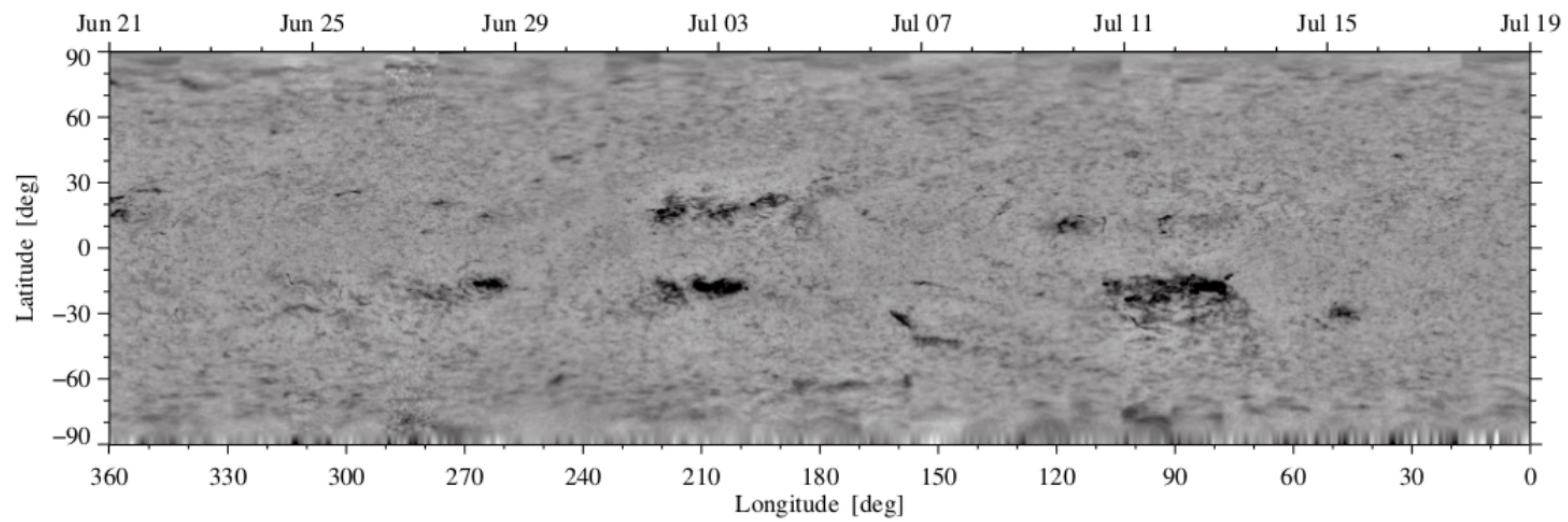}
   \caption{Carrington map (rotation No.\ 2125) derived from calibrated ChroTel 
\mbox{He\,\textsc{i}} line-core images covering the time period from 2012 
June~21 to end of July~18. The Carrington map comprises one calibrated full-disk 
image per day, which was derotated to the respective Carrington latitude with a 
sampling of 0.1$^\circ$.} 
   \label{Fig:carrington}%
    \end{figure*}

The reference map obtained with GRIS is dominated by stronger red-shifted 
profiles, representing down-streaming flows in the upper part of the Doppler map 
(Fig.~{\ref{Fig:dop3}\ns}a). In the lower part, weaker blue-shifted profiles 
are encountered, which represent up-streaming flows. The spatial resolution of 
GRIS is 0.13\arcsec~pixel$^{-1}$, i.e., roughly ten times better than the 
ChroTel resolution. As expected from the low spatial and spectral resolution, 
the ChroTel maps lack fine structure compared to the GRIS map and thus appear 
pixelated in a direct comparison. More importantly, for each ChroTel velocity 
value, about 100 GRIS measurements are available, representing different 
physical conditions in the chromosphere. In consequence, a very tight 
correlation of the ChroTel and GRIS velocity measurements is not expected. 
However, the overall structure is recovered in the ChroTel maps, especially in 
regions with large velocities. Furthermore, the two ChroTel Doppler maps in 
Fig.~{\ref{Fig:dop3}\ns} are very similar. Next, we will analyze scatter plots 
of GRIS and ChroTel Doppler velocities for both methods to validate the 
velocity values in the ChroTel maps.  

Figure~{\ref{Fig:gris_vch}\ns} shows the velocities from the difference map 
plotted against the reference velocities from the GRIS Doppler map. In the 
original calibration runs, \citet{Bethge2011} found that reconstructed 
velocities fall into two regimes: $-5 < v < 5$~km~s$^{-1}$ and $|v| \ge 
5$\,km~s$^{-1}$. In our analysis, we find that a similar separation occurs at 
6\,km~s$^{-1}$. Thus, the following scatter plots will focus on the velocity 
regime $-6 < v < 6$~km~s$^{-1}$. From the best linear fit, overplotted in gray, 
we obtain a slope of m = 0.0031\,s~km$^{-1}$ and a vertical intercept of 0.0015. 
The factor $C$ determines to $C = 1/m = 320$\,km~s$^{-1}$, which is used to 
calculate Doppler velocities from the difference map. Linear fits to smaller 
velocity intervals show the same slope, which indicates that a linear model is 
appropriate for this relationship. Pearson's linear correlation coefficient 
between the reconstructed and reference velocities in this regime is $\rho = 
0.48$. In comparison, \citet{Bethge2011} found a correlation coefficient of 
$\rho =0.45$ using the central three filters in the $-5 < v < 5$~km~s$^{-1}$ 
regime. Thus, a simple difference map with the Zernike-calibrated filtergrams 
yields comparable results, without resorting to external calibration.  

Figure~{\ref{Fig:gris_vlos}\ns} shows the reconstructed velocity from the 
parabolic fit vs.\ the GRIS velocity in the $-6 < v < 6 $~km~s$^{-1}$  regime, 
and the linear fit with a slope of 0.51. The vertical intercept of 0.70 reflects 
a difference in the rest wavelength. As seen in the scatter plot, the parabolic 
fit velocities show a larger scatter around the linear trend line, and the 
correlation coefficient is only $\rho =0.27$. In both scatter plots, the 
horizontal striation is a result of enlarging ChroTel data to match the 
resolution of GRIS. Each pixel in ChroTel data covers many pixels in GRIS data 
and consequently a range of GRIS velocities. As a sanity check, we computed the 
linear correlation between the two types of ChroTel Doppler velocities. The high 
correlation of $\rho = 0.77$ suggests that these two methods deliver consistent 
results. Finally, a full-disk Doppler map is displayed in 
Fig.~{\ref{dop_disk_3p_plot}\ns} as an example, which corresponds to the best 
set of filtergrams on 2014 May~12.


\section{Discussion and conclusions}

The difference method is more suitable for producing full-disk maps than the 
parabolic fit method. In regions with low or no \mbox{He\,\textsc{i}} 
absorption, spectral line fitting is not an option. In our sample data sets, the 
parabolic fit is only possible in regions with strong \mbox{He\,\textsc{i}} 
absorption, i.e.,  active regions and filaments. Thus, full-disk Doppler maps 
are produced with the difference method. We present the parabolic fits for 
active-region data mainly as a cross check for the consistency of the results. 
Compared to the COG method provided in the original work of \citet{Bethge2011}, 
the difference map from Zernike-calibrated filtergrams does not require each 
time an external calibration with a co-temporal spectrograph scan.  

Re-calibrating the entire ChroTel archive requires considerable computational 
resources.  In the Zernike calibration procedure, the most computationally 
intensive part is performing a singular value decomposition on each image. 
Therefore, modeling the long-term temporal evolution of the Zernike coefficients 
could minimize computation time. It is possible to avoid repeated fitting 
procedures if the daily and annual variations of the coeffients can be modeled.
 
With this new method, ChroTel dopplergrams will no longer be limited by the 
availability of high-resolution spectrographic data to calibrate the data. Using 
this method, we are able to produce synoptic maps with ChroTel 
\mbox{He\,\textsc{i}} filtergrams as shown in Fig.~{\ref{Fig:carrington}\ns}. 
It uniformly re-normalizes all filtergrams across different days. Synoptic 
ChroTel data allows studies of solar cycle variations and for solar activity 
monitoring. Solar features can be identified with digital image processing in 
ChroTel filtergrams. The same method can be employed to reduce the H$\alpha$ 
and \mbox{Ca\,\textsc{ii}}\,K data from ChroTel. In addition, the 
high-resolution observations of GRIS and other high-resolution spectrographs can 
be embedded in the large-scale context with full-disk \mbox{He\,\textsc{i}} 
filtergrams and Doppler maps. The ChroTel data complements SDO data with data in 
three chromospheric wavelength.

The calibration method developed in this paper can potentially be applied to 
other full-disk images. As an example, images from the the Narrowband Filter 
Imager (NFI) on board the Hinode Solar Optical Telescope \citep{Tsuneta2008} 
contain artifacts which are caused by air bubbles in the fluid inside the 
tunable filter. They distort and move when the filter is tuned, and then usually 
drift toward the edges of the field of view over time \citep{Tsuneta2008}. 
Zernike polynomials may be appropriate to model these image artifacts and remove 
them. Similarly, the Zernike method can be used to characterize background 
trends and optical artifacts in other solar full-disk images.


\section*{Acknowledgments}
ChroTel is operated by the Kiepenheuer Institute for Solar Physics (KIS) in 
Freiburg, Germany, at the Spanish Observatorio del Teide on Tenerife (Spain). 
The ChroTel filtergraph was developed by the Kiepenheuer Institute in 
cooperation with the High Altitude Observatory (HAO) in Boulder, Colorado. ZS's 
research internship in Germany was made possible by the Research Internships in 
Science and Engineering (RISE) program of the German Academic Exchange Service 
(DAAD). CD acknowledges support by grant DE 787/5-1 of the 
German Research Foundation (DFG). The authors thank Drs. C. 
Bethge and W. Schmidt for carefully reading the manuscript and helpful comments.

%

\end{document}